\documentclass[aps,jcp,reprint,superscriptaddress]{revtex4-2}

\usepackage{graphicx}
\usepackage{eqnarray,amsmath}
\usepackage{dcolumn}
\usepackage{bm}
\usepackage{amssymb}
\usepackage{alphabeta}
\usepackage{comment}
\usepackage[nolist, nohyperlinks]{acronym}
\usepackage[mathlines]{lineno}
\usepackage[dvipsnames]{xcolor}
\usepackage[version=4]{mhchem}
\usepackage{appendix}
\usepackage{hyperref}
\usepackage{cleveref}

\begin{document}%
\title{Dissipative split-charge formalism: Ohm's law, Nyquist noise, and non-contact friction}

\author{Martin H. Müser}
\affiliation{Dept. of Materials Science and Engineering, Saarland University, Germany}

\date{\today}

\begin{abstract}
The split-charge equilibration method is extended to describe dissipative charge transfer similarly as the Drude model, whereby the complex-valued frequency-dependent dielectric permitivities or conductivities of dielectrics and metals can be mimicked at non-zero frequencies. 
To demonstrate its feasibility, a resistor-capacitor circuit is simulated using an all-atom representation for resistor and capacitor. 
The dynamics reproduce the expected charging process and Nyquist noise, the latter resulting from the thermal voltages acting on individual split charges.
The method bears promise to model friction caused by the motion of charged particles past metallic or highly polarizable media. 
\end{abstract}

\date{\today}
\maketitle

\section{Introduction}

The electronegativity equalization method~\cite{Nalewajski1984JACS,Mortier1985JACS,Mortier1986JACS} and its charge-equilibration (QE)~\cite{Rappe1991JPC} or fluctuating-charge~\cite{Rick1994JCP} variants are frequently used in molecular simulations to assign partial charges on the fly.
It can be seen as a low-key coarse-grained approximation of density-functional theory, where spatial discretization points coincide with atomic positions~\cite{Nalewajski1984JACS,Mortier1985JACS,Mortier1986JACS,Verstraelen2013JCP}. 
In the simplest case, the total energy of a system is a quadratic polynomial in the set of partial charges $\{Q_i\}$~\cite{Muser2022APX,Jensen2023JCTC}, see also Eq.~\eqref{eq:define_QE}.
The zero-order coefficient generally depends strongly on the environment and contains, e.g., pair repulsion and many-body cohesion as described in the embedded-atom-method~\cite{Streitz1994PRB}.
The first-order term is, depending on background or application, the electronegativity of an isolated atom~\cite{Nalewajski1984JACS,Mortier1985JACS,Mortier1986JACS} or the work function of a metal~\cite{Muser2012EPJB}, while the second-order term represents the self-interaction of the charge plus Coulomb interactions between atoms. 
It can be parameterized to reflect either the chemical hardness of an isolated atom~\cite{Mortier1985JACS,Mortier1986JACS} or the Thomas-Fermi screening length in a metal~\cite{Nistor2009PRB,Scalfi2020JCP}. 
Adding enviroment dependence to the QE parameters~\cite{Bhattarai2019PRB} promises to increase the transferability, such that properties of both isolated atoms and solids but potentially also any intermediate structure can be faithfully reproduced.
However, even if successful, regular QE schemes still face various intrinsic limitations.
The two most relevant to this work are that
any QE solid has the static dielectric response function of a metal~\cite{Nistor2009PRB}.
Moreover, QE is a pure equilibrium approach lacking dissipation caused by time dependence. 

The metallicity problem of regular QE approaches could be solved with the split-charge equilibration (SQE) method~\cite{Nistor2006JCP,Nistor2009PRB,Verstraelen2009JCP}.
It arose as a phenomenological split between the conventional QE and a pure bond-polarizibility model~\cite{Chelli1999JCP}.
In brief, SQE allows transfer of fractional charges only through chemical (covalent and metallic) bonds.
In non-metallic systems, this transfer is penalized through a bond hardness term in addition to that caused by the chemical or atomic hardness.
The latter can also be losely associated with the Anderson $U$ parameter describing the strength of the on-site charge interaction in the Hubbard model~\cite{Hubbard1963PRSLA}.
The dielectric permittivity of a solid turns out inversely proportional to the bond hardness~\cite{Nistor2009PRB}.
In the redox-SQE formalism, in which integer charges can be donated from one atom or ion to another one without bond penalty, (irreversible, contact-induced) electron-transfer dielectrics can be simulated~\cite{Dapp2013EPJB}. 
This feature made it possible to simulate battery discharge and recharge, such that  all processes inside the battery were modelled atomistically, including the build-up of voltage between the electrodes~\cite{Dapp2013JCP}. 
A 'daemon resistor', which was not modelled explicitly, allowed current to flow externally between the electrodes according to Ohm's law.
%
%
%
In this study, the roles of all-atom and daemon swap:
the battery is the daemon, while the electrical circuit component is modeled explicitly. 

To achieve a self-containing all-atom representation of resistors and capacitors, we add an inertial and, more importantly, a dissipative (D) term to the SQE formalism.
The resulting D-SQE approach, 
which can be seen as a hybrid of the Drude  model, 
the polarizable-ion model~\cite{Lewis1985JPCSSP,Cipcigan2019RMP},
and a related extension of polarizable force fields~\cite{Cheng2022JCP}
can mimick frequency-dependent conductivities of metals or dielectric permitivitties of dielectrics over a broad frequency range and not merely their quasi-static limits.
Moreover, resistors can be put together using (coarse-grained) atomic building blocks and when subjected to thermal random forces, or rather random voltages, Nyquist noise in capacitors arise naturally.

When broken down to a truly atomic scale, for example, to a hypothetical one-atom-thin wire, the model can certainly not (yet?) reproduce the intricacies of Luttinger liquids~\cite{Haldane1981JPCSSP}. 
The goal of the current work is to design simple microscopic charge-transfer dynamics that produce the appropriate response functions of bulk systems.
%
Despite its simplicity, the pursued approach can be expected to further improve the reliability of simulations of ions near electrodes~\cite{AhrensIwers2022JCP,Goloviznina2024AM,Holoviak2024JPCA} or related systems, where charged particles move relative to metallic or highly polarizable surfaces.
Thus, this work should be viewed as a feasibility test rather than as an attempt to describe a specific solid. 
To this end, the model is introduced first. 
Next, its generic dielectric response functions are derived. 
Following that, simulations of an RC element driven by either a battery or thermal noise or by a fixed charge moving through the capacitance are presented. 
Finally, conclusions are drawn.

\section{Model and theory}

\subsection{General model}
\label{sec:general_model}

In this study, all grid-points or (pseudo-) atoms participating in charge-transfer
are fixed in space. 
This allows us to ignore the zero-oder term $U_0(\{r_{ij}\})$, which can be argued to reflect the energy of a system under the constraint that all atoms be neutral, or, rather to have an integer oxidation state. 
This  'expansion coefficient' would surely depend on the method used to assign partial charges.
To proceed, let us make the unplausible but pragmatic assumption, that we identified one such method approved by readers, referees, and author alike. 
The relevant, remaining charge-transfer energy then reads 
\begin{equation}
\label{eq:define_QE}
\Delta U = 
- \sum_i \chi_i Q_i 
+ \frac{1}{2} \sum_{i,j} \left\{ J_{ij}(r_{ij}) Q_i Q_j + \kappa_{ij}q_{ij}^2  \right\}
\end{equation}
in the split-charge formalism, where 
\begin{equation}
Q_i =  n_i e + \sum_j q_{ij}
\label{eq:defintion_sq_charge}
\end{equation}
is the charge of atom $i$, while $q_{ij} = -q_{ji}$ is the partial charge or split charge donated from atom $j$ to atom $i$, which is penalized via a bond-hardness term $\kappa_{ij}$. 
For reasons of simplicity, we assume neutral rather than ionic references and thus omit the oxidation state $n_i$ in the following. 
The $\chi_i$ denote the electronegativity of atom $i$ to which a potential bias or the interaction with external charges can be added.
The diagonal elements $J_{ii}$ represent the self-interaction or (half) the chemical or atomic hardness of atom $i$.
In the following, mono-atomic model, they will be denoted as $\kappa_\textrm{a}$. 
The off-diagonal elements $J_{i,j\ne i}$ represent the Coulomb interactions between atoms.
They can be damped at short distance to mirror the delocalization of charge density~\cite{Rappe1991JPC}.
Here, they will simply be set to $J_{i,j\ne i} = 1/r_{ij}$, where $r_{ij}$ is the distance between atoms $i$ and $j$. 

While split charges are not needed in the QE formalism for the modeling of (singly connected) metals, in which $\Delta U$ is merely minimized w.r.t. the set of charges $\{Q\}$, the locality of charge transfer must be imitated whenever electron dynamics matter.
To reflect locality, $\kappa_{ij}$ is assumed to be zero between nearest neighbors in a metal and approximated as infinitely large otherwhise to make the value and thereby the energy of that split charge vanish.
In this zero-order approach, split charges only need to be considered between nearest neighbors.

The driving force or voltage acting on a split charge is $V_{ij} = -\partial \Delta U/\partial q_{ij}$, i.e., when using $\partial Q_k\partial q_{ij} =  \delta_{ik} - \delta{jk}$, 
\begin{equation}
V_{ij} = - J_{ij}(Q_i-Q_j) - \kappa_{ij}q_{ij} + (\chi_i-\chi_j).
\end{equation}
%
%
To model dissipative dynamics \textit{locally}, rather than globally as, e.g., through Eq.~(3) in Ref.~\cite{Holoviak2024JPCA}, 
\begin{equation}
\label{eq:eq_of_motion}
L_{ij} \ddot{q}_{ij} + R_{ij} \dot{q}_{ij} + \kappa_{ij} q_{ij} = - \Phi_i + \Phi_j  + \chi_i-\chi_j + V_\textrm{th}(t)
\end{equation}
is proposed as equation of motion, where $\Phi_i = \sum_j J_{ij} Q_j$ is the electrostatic potential of atom $i$ produced by the explicitly treated charges. 
$L_{ij}$ is a pseudo inductance arising due to electronic inertia, while $R_{ij}$ absorbs dissipative effects in a phenomenological fashion.
This generalization is similar in spirit to the (Lorentz-) Drude model and will thus share many strentghs and limitations.
At finite temperature, dissipation entails thermal noise $V_\textrm{th}(t)$, 
whose expectation values must satisfy
\begin{subequations}
\begin{eqnarray}
\langle V_\textrm{th}(t) \rangle & = & 0 \\
\langle V_\textrm{th}(t) V_\textrm{th}(t') \rangle & = &  2 k_B T R \delta(t-t')
\end{eqnarray}
\end{subequations}
for an instantaneous Ohmic voltage of $R \dot{q}$,
according to the fluctuation-dissipation theorem~\cite{Kubo1966RPP}.
Here $k_B T$ is the thermal energy and $\delta(t-t')$ is Dirac's delta function. 
%
%

Connecting the resulting dissipative SQE model and the Drude model is most easily done by discretizing a metal into cubic elements, as this avoids the need for book keeping of Cartesian indices. 
Alternatively, let us consider polononium, which is the only element condensing into a simple cubic (sc) lattice. 
An instantaneous current of an elementary charge $e$ per time $\tau$ between each two nearest neighbors along one principle direction implies a current density of $j = e \rho v \to e (1/a_0^3) (a_0/\tau)$ in the Drude model, where $a_0$ is the bond length. 
Replacing $e/\tau$ with $\dot{q}$ means that the SQE current density is $\dot{q}/a_0^2$.
Ohm's law is thus satisfied correctly if the split-charge resistance is parameterized as $R = \rho a_0$, where $\rho$ is the resistivity. 
Similarly, the kinetic energy $m v^2/2$, $m$ being the (effective) electron mass turns into $L \dot{q}^2/2$ with  $L = m a_0^2 / e^2$. 
The numerical value of $L$ can be assigned to target the plasma frequency.
%
Before addressing this and other dielectric properties in more detail, the default unit system and useful dimensionless constants are introduced. 

The units for mass $[m]$, charge $[Q]$, and length $[l]$ are the electron mass $m_\textrm{e}$, $e$, and $a_0$, respectively.
The temperature $T$ is expressed as thermal energy, while $[E] = e^2/(4\pi\varepsilon a_0)$ is the unit of energy.
%
%
Occasionally, we will revert to extended SI units.
This will be marked by an upper index SI on the r.h.s. of pertinent equations. 
Otherwhise, we use reduced units.
For sc $\alpha$-Po, non-standard base units are $a_0 = 3.345$~{\AA}, $[E] = 4.305$~eV, and $[T] = 49,960$~K, while selected derived units are $[t] = 0.3842$~fs, resistivity $[\rho] = 3.455~\mu\Omega\cdot$m, and $L = 3.971$~pF.

A central number of a mono-atomic SQE model is the dimensionless chemical hardness $J_{ii}$.
Its value for polonium can be estimated from its
electron affinity $A \approx 8.42~\text{eV}$ and its ionization energy $I \approx 1.9~\text{eV}$ to be $J_{ii} = A + I \approx 10.3$~eV/$e^2$, i.e., $\approx 2.4$ in reduced units.
This exceeds the Madelung constant of the rocksalt structure (rs), $\alpha_\text{M}= 1.7476$.
Thus, even when using undamped Coulomb interactions, a spontaneous symmetry breaking from sc to rs is energetically unfavorable.
Since rs is the softest charge-transfer direction for the sc lattice~\cite{Muser2012EPJB}, the pursued simple parameterization makes the energy of a finite system be positive definite in the split charges. 
 

\subsection{Frequency-dependent dielectric constant}
\label{sec:dielectric_constant}

In former work~\cite{Nistor2009PRB}, it was shown that the wave-number-dependent, static dielectric permittivity of the SQE model on the simple-cubic lattice is given by
\begin{equation}
\label{eq:static_permittivity}
\tilde{\varepsilon}_\textrm{r}^\text{SI}(\mathbf{k}) = 1 + \frac{1}{\varepsilon_0 a_0(\kappa_\textrm{b} + \kappa_\textrm{a} k^2 a_0^2)}
\end{equation}
in the SI unit system.
Here, $\kappa_\textrm{b}$ is the split-charge or bond (b) hardness, which replaces the nearest-neighbor $\kappa_{ij}$. 
Since the diverging next-nearest $\kappa_{ij}$ lead to (infinitesimally) small split-charges and bond-charge energies, their effect can be ignored.
The validity of Eq.~\eqref{eq:static_permittivity} hinges on various approximations, such as undamped Coulomb interactions and environment-independent $\kappa_\text{a}$, which, however can be accounted for in principle. 
In fact, simple-cubic-lattice specific discretization corrections to the continuum Coulomb (C) interaction, $\tilde{J}_\textrm{C}(\mathbf{k}) \propto 1/k^2$, have proved important to reproduce the Thomas-Fermi screening length $\zeta$~\cite{Nistor2009PRB}. 
It is also noted that $\kappa_\textrm{b}$ is supposedly the most difficult but also the most important coefficient to be made environment dependent, as it will depend not only on the distance but also on local coordination numbers or bond order. 

We revert to the unit system of this study by replacing $\varepsilon_0$ with $1/4\pi$. 
Since all equations are linear, the static bond hardness $\kappa_\textrm{b}(\omega = 0)$ only needs to be replaced with $\kappa_\textrm{b}(\omega) = -\omega^2L - \text{i}\omega R + \kappa_\textrm{b}(0)$ to yield
%
\begin{equation}
\label{eq:dynamic_permittivity}
\tilde{\varepsilon}_\textrm{r}(\mathbf{k},\omega) = 
1 + 
\frac{4\pi/a_0}{-L\omega^2-\text{i}R \omega + \kappa_\text{b} + \kappa_\text{a}k^2a^2}.
\end{equation}
For a loss-free metal, $R = \kappa_\text{b} = 0$, it follows that the dielectric permittivity obeys
$\epsilon_\textrm{r} = 1 - \omega^2_\textrm{p}/\omega^2$
at scales large compared to atomic spacings ($k \to 0$), where $\omega_\textrm{p} = 4\pi/(a_0L)$ is the plasma frequency. 
Thus, when used to model the dielectric properties of a metal ($\kappa_\text{b} = 0$)  using a simple-cubic discretization, $L$ can be parameterized to reproduce the plasma frequency.

A second resonance frequency could be realized, for example, by introducing next-nearest split charges. 
Another generalization would be to account for atomic polarizability.
In other words, the D-SQE formalism
allows for a quite flexible parameterization of  $\tilde{\varepsilon_0}(\mathbf{k},\omega)$.
However, the flexibility already contained in Eq.~\eqref{eq:dynamic_permittivity} should suffice for most practical applications. 

To overcome the limitation to sc structures, the derivation of Eq.~\eqref{eq:static_permittivity} must be repeated for other (Bravais) lattices.
This leads to a rescaling of the parameters  $L$, $R$, $\kappa_\textrm{s}$, and $\kappa_\text{a}$ in Eq.~\eqref{eq:dynamic_permittivity} with $2D/Z$ and/or with $a_0^D/V_\text{ec}$, where $Z$ is the coordination number, $D$ the spatial dimension, and $V_\textrm{ec}$ the volume of the elementary cell. 
However, for Eq.~\eqref{eq:dynamic_permittivity} to be accurate
damping and continuum corrections may have to be applied.
Exploring this in more detail is beyond the scope of the present feasibility test with its focuses on a generic demonstrator, which is discussed next.

\subsection{Demonstrator model}
\label{sec:demonstrator}

\begin{figure}[hbtp]
\includegraphics[width=0.485\textwidth]{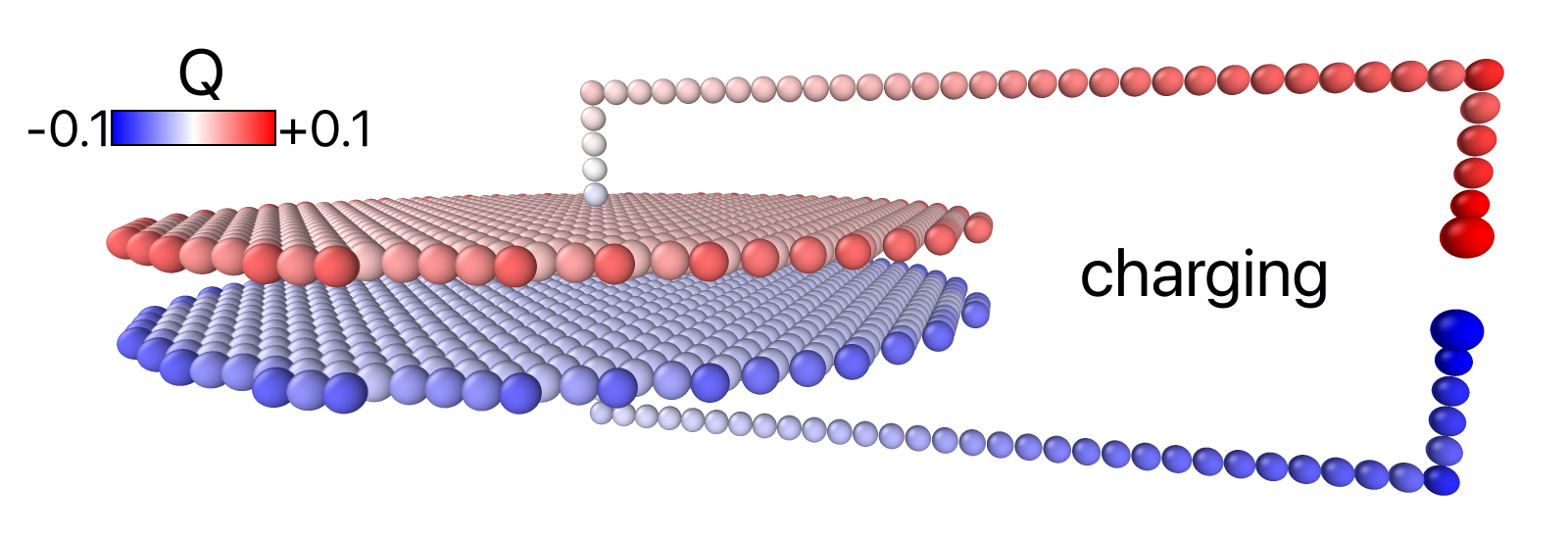}
\includegraphics[width=0.485\textwidth]{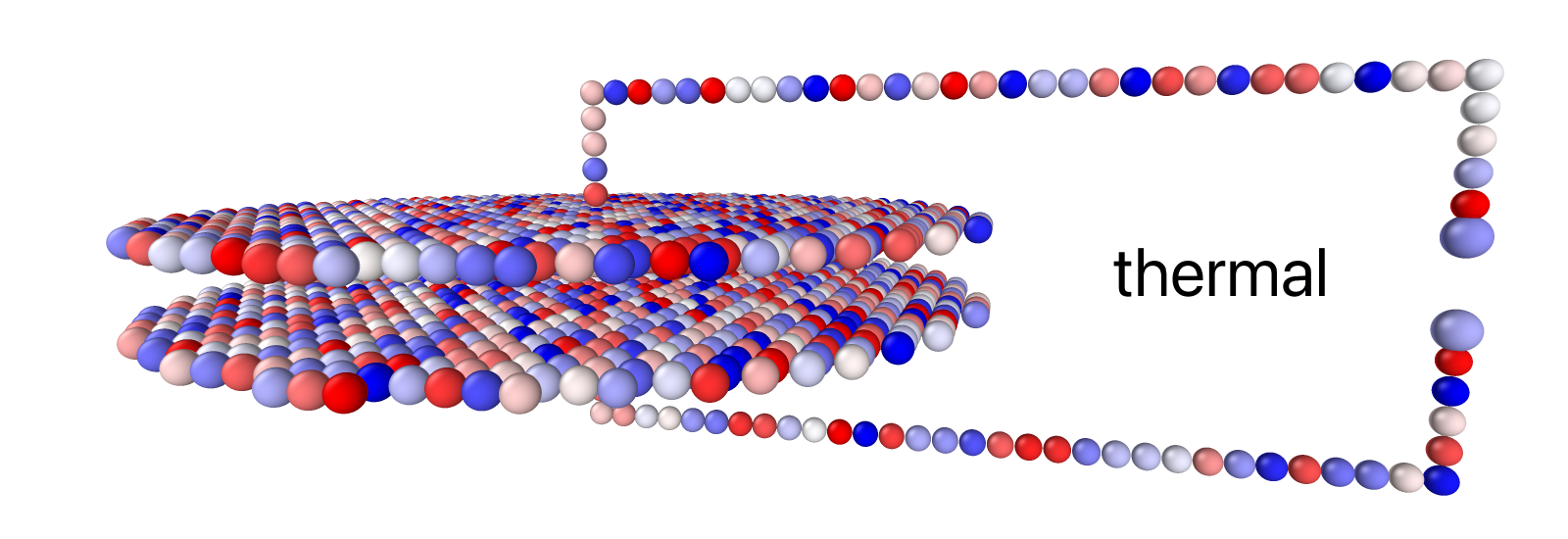}
\includegraphics[width=0.485\textwidth]{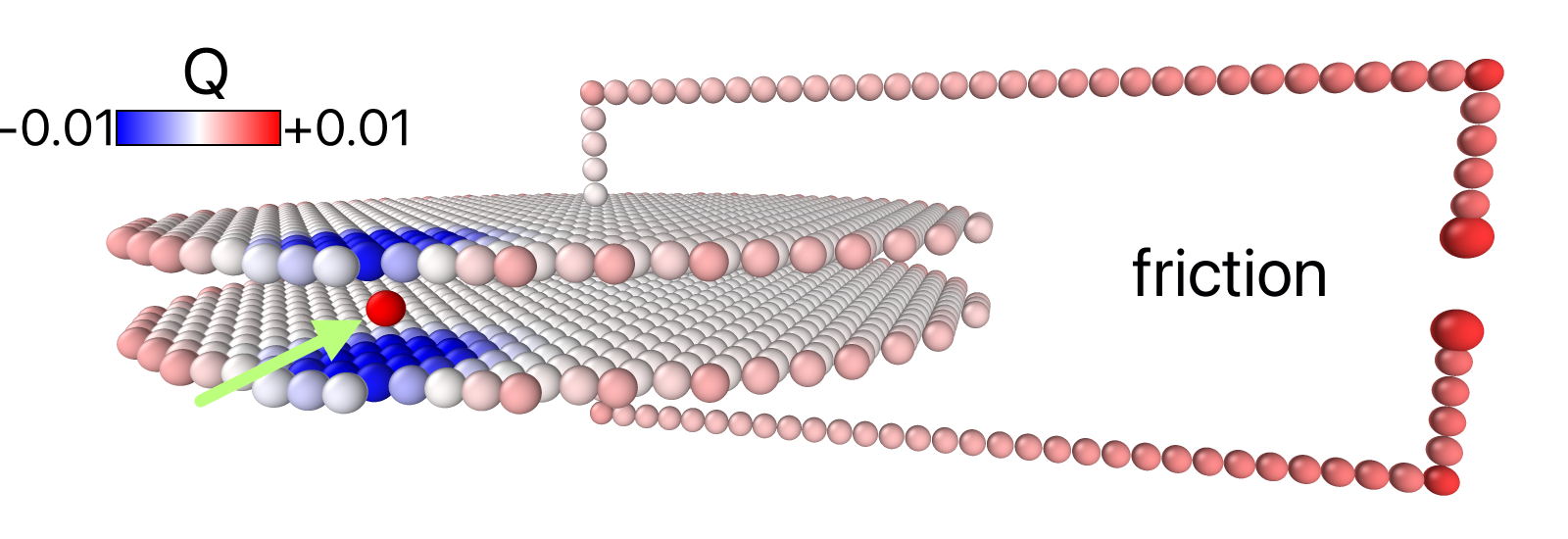}
\caption{\label{fig:capaModel}
Set-up of the model in three modes of operation.
Spheres represent atoms, with colors indicating charges.
Split charges connect adjacent atoms and wire ends.
%
}
\end{figure}

The D-SQE model's ability to capture charge-transfer dynamics is examined using a generic, all-atom resistor-capacitor (RC) setup.
The RC element is coupled to an external voltage, thermal fluctuations causing Nyquist noise, or a fixed charge passing through the capacitor, as shown in Fig.~\ref{fig:capaModel}.

All atoms and bonds of the RC element are treated equally.
%
Parameters are given without units; however, their values are inspired by those describing polonium.
This choice is not significant, as the unit system could be based on mesoscopic rather than atomistic discretization elements.
Specifically, the nearest-neighbor spacing is $a_0 = 1$, while
the atomic hardnesses is set to $\kappa_\textrm{a} = 2.4$.
Atoms have a zero electronegativity, except for the two terminal atoms, for which we set $\chi = \pm V/2$ when applying a voltage $V$. 
%
All split-charges are given a unity pseudo inductance, for the author's inability to find a literature value for polonium's plasma frequency.
Any split-charge resistance
is set to $R_\text{sq} = 0.1245$. 

A capacitor plate has a target radius of $r \approx 15a_0$ resulting in $N_\text{p} = 717$ atoms per plate. 
The two plates are separated by a distance of $d = 3a_0$, yielding a continuum-theory capacitance of $C = N_\text{p}a_0^2/(4\pi d) \to 19$.
A better value for $C$ is obtained by equally distributing a total unit charge of opposite sign on the two plates and computing the energy, which gives $C = 26.4$.
Further corrections have opposite signs.
If factors related to atomic hardness were included, the estimated capacitances would decrease, while accounting for the flexibility of some (excess) charge densities on the wires and the plates' rim would have the opposite effect.

Each wire strand consists of 39 atoms leading to a total external resistance of $R = (2\times 39+1) R_\text{sq}$ given the extra split charge between wire terminals. 
However, Ohmic dissipation also occurs in the capacitance, even if current flows in a plane rather than through a wire. 
In order to reduce the relative importance of that contribution to the overall resistance, split charges were also placed between next-nearest neighbors.
Note that this eliminates (a little more than) one split-charge resistance at each of the four wire corners as well as that of the split-charge connecting the wire ends to the plates.
An improved guess for the total resistance thus is $R = 73~R_\textrm{sq} \to 9.1$. 

%
A few words on the split charge between the wire ends may be appropriate.
As long as no voltage is applied between these two 'atoms,' their bond is simply longer but has the same resistance as any other bond.
Turning the wire ends into (daemon) battery terminals requires adding the voltage $V$ to the battery split charge, which, however, now represents ionic rather than electronic charge donation. 
(This added voltage is the critical factor rather than assigning electronegativities of $\pm V/2$ to the two terminal atoms.) 
While it would have been physically meaningful to assign a different, potentially much larger value to the battery's internal resistance than to regular bonds, the original value was kept primarily so that the circuit's resistance did not depend on the specific mode of operation. 
Similar comments applies to the hardness of the terminal atoms. 

Before starting the charging simulation, the system is equilibrated with an open switch, preventing (ionic) charge transfer between the battery terminals at 'negative times.'
This results in a marginal charge transfer between each electrode and the half RC element it is connected to.

The Python code developped to simulate the demonstrator model is avaiblable under \url{https://github.com/mueser/D-SQE}.
The charging simulations of 200\,000 time steps took about 92~s on a two-year old MacBook Pro when atoms with a nearest-neighbor distance were connected through a total of $N_\text{sq} = 2\,825$ split charges. 
These numbers changed to 108~s and $N_\text{sq} = 5{,}533$ when also next-nearest neighbors were connected.
The 15\% increase of total computing time, after $N_\text{sq}$ was essentially doubled, implies that the excess computing time related to split charges was merely about 30\%.
Thus, the advantage of avoiding explicit split charges while reproducing proper dissociation limits with a charge transfer model (QTM), such as the one proposed by Gergs~\cite{Gergs2021JCTC}, is primarily related to memory, not computation speed.
Despite its elegance, it remains unclear how to incorporate a targeted dielectric response function into QTM.

\section{Results}

Before formally discussing the results in this section, we encourage the reader to view the movies M1--M5 uploaded to the electronic supplement (and for the time before publication being available at \url{https://github.com/mueser/D-SQE}), as this may facilitate the understanding of the text.
M1 illustrates the initial stages of the charging process of the RC element. 
M2 extends the observation over a longer time span but with lower temporal resolution, while M3 captures the oscillatory charging dynamics when the split-charge resistivity is set to zero.
M4 highlights the dynamics under thermal noise.
Finally, M~5 shows the effects in the presence of a fixed charge passing through the capacitor.

\begin{figure}[t!htbp]
\vspace*{5mm}
\includegraphics[width=0.475\textwidth]{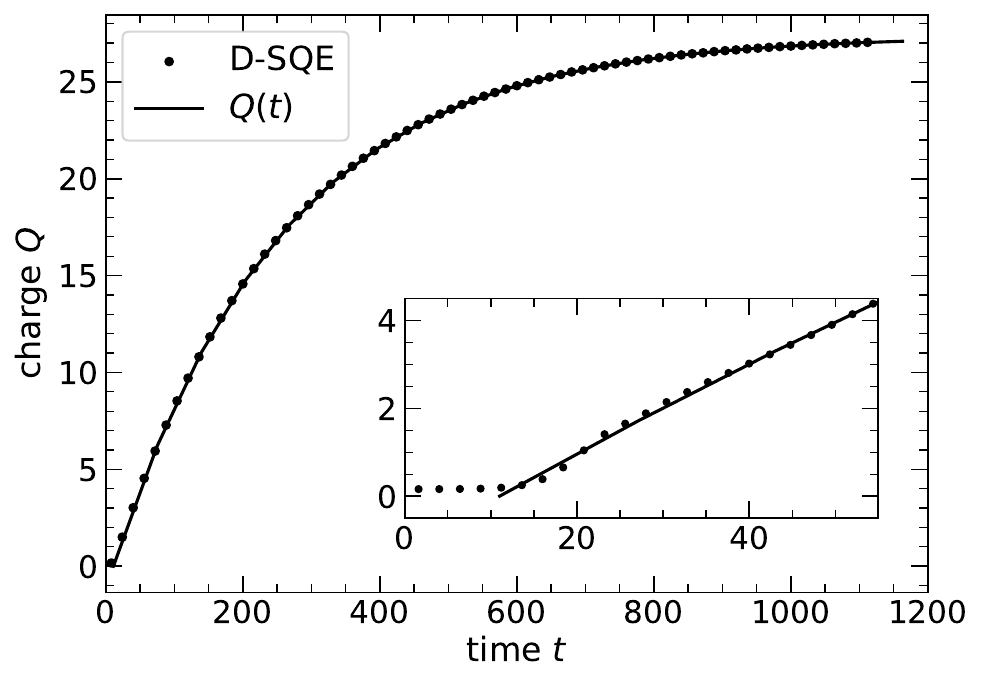}
\vspace*{-2mm}
\caption{\label{fig:charging}
Time dependence of the charge $Q$ as obtained in the D-SQE model (symbols) and according to the solution of $Q(t)$ (lines), see Eq.~\eqref{eq:solution_ODE}.
The inset highlights the early-time behavior. 
}
\end{figure}

The charging of the capacitor at zero temperature is addressed first. 
As a reminder, the ordinary differential equation describing the charging procees of a macroscopic RC element coupled to a direct current battery is 
\begin{equation}
R \dot{Q} + \frac{1}{C} Q = V \Theta(t-t_0).
\end{equation}
Its solution for the initial condition $Q(t=t_0) = 0$ is
\begin{equation}
Q(t) = C V \{1 -  \exp(-(t-t_0)/\tau)\} \Theta(t-t_0),
\label{eq:solution_ODE}
\end{equation}
with a relaxation time of $\tau = RC$.
Fig.~\ref{fig:charging} shows that this equation fits the data quite well.
Moreover, the  asymptotic charge of $Q = 27.36$ is surprisingly close to the value of $Q = 26.4$, which was 'predicted' in Sect.~\ref{sec:demonstrator}.
Since the voltage is unity, capacitance and equilibrium charge assume identical numerical values. 
From the relaxation time $\tau = 248.7$, an effective resistance of $R = \tau/C \to 9.09$ can be deduced.
This corresponds almost exactly to the simple estimate of $R = 9.1$, which was also made in Sect.~\ref{sec:demonstrator}.
Such a close agreement must benefit to some degree from fortuitous error cancelation:
The resistance of each corner was slightly over- and that of the capacitor underesimtated, each time by an amount, which certainly exceeds the 0.1\% deviation between the simulation result and the back-of-the-enveloppe estimate. 
Here, and in the following, we would deem deviations between simulation results and expectations as acceptable, whose order does not exceed $a_0/r \to 0.067$, since this ratio gives the order of magnitude of either the number of wire or rim atoms relative to the number of atoms per plate. 

Two further observations are worth discussing.
First, the electrical circuit is completed at time $t = 0$.
However, the charge on the capacitor's plates starts increasing noticeably only at $t_0 \approx 11.0$. 
This delay results from the time it takes for the information to travel from the 'switch,' in this case the battery, to the front atom of the capacitance. 
Since the calculations for the dielectric permittivity does not transfer directly to the one-dimensional wire geometry, we do not know the precise wave speed $c$ with which the split-charge travels through the wire.
However, it is plausible that it is of the order of $\sqrt{\kappa_\textrm{a}a/L} \to 1.55$.

A second interesting observation is that the plate connected to the cathode is marginally positive before the circuit is completed.
This occurs because the cathode has a higher electronegativity, attracting 'electrons' to it even before charge-neutralizing 'ions' can flow between the two terminal atoms.
Once the switch is closed, charge compensation occurs.
Current flows until the chemical potential of the atomic charges on the two opposing plates differs by the voltage applied to the wire/battery terminals.

\begin{figure}[t!htbp]
\includegraphics[width=0.475\textwidth]{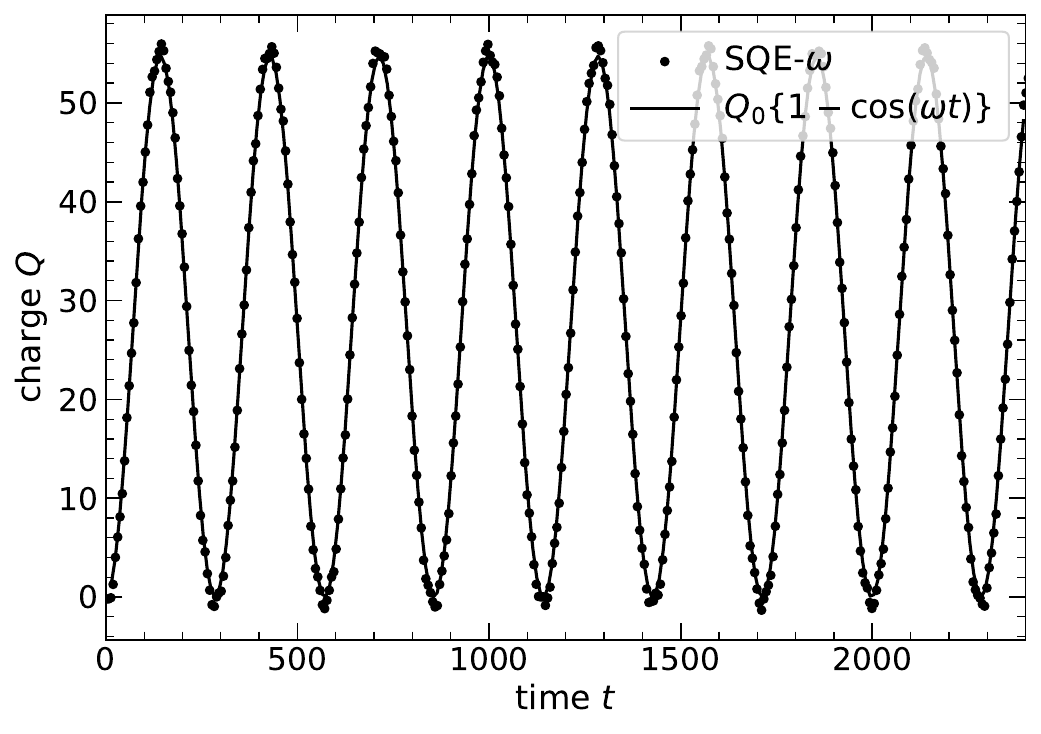}
\caption{\label{fig:charging_2}
Time dependence of the charge $Q$ as obtained in the SQE-$\omega$ model (symbols) and according to the formal solution $Q(t) = Q_0\{1-\cos(\omega t)\}$.
}
\end{figure}

The D-SQE model no longer dissipates energy when the split-charge resistance $R_\text{sq}$ is set to zero.
In this case, the model could be called the SQE-$\omega$ method, in analogy to the so-called ACKS2$\omega$ model~\cite{Cheng2022JCP}, which motivates an on-site frequency dependent SQE model from the bottom up rather than in a top-down fashion. 
(To complete the isomorphism to the ACKS2 description, point dipoles and potentially higher-order multipoles would have to be added to the SQE description, while the resistance would have to be set to zero.)
Omitting the resistance changes the nature of the circuit from an RC element to an LC element, where L stands for inductor.
In the given case, its inductance $L$ would be a collective pseudo inductance, which has a similar effect on the dynamics as a magentic inductance. Its value would be close to the sum of the inertia of individual split charges.
The corresponding oscillations of the LC element are depicted in Fig.~\ref{fig:charging_2} for the same initial condition as for the RC circuit. 
Small deviations from a single-sinusoidal oscillations are noticeable.
They are due to a coupling of the capacitor's (symmetrized) charge to eigenmodes other than the slowest one, even if the slowest eigenmode clearly dominates the signal for the given initial boundary condition. 

%
%

The Nyquist noise was also investigated.
To this end, the same setup was used as before, except that both wire end atoms were assigned the same chemical potential as the other atoms, and no voltage was applied.
The measured observable is the time auto-correlation function (ACF) of the capacitor's charge $Q$, from which the noise spectrum can be deduced via a Fourier transform:
\begin{equation}
\label{eq:acf_q}
C_{QQ}(\Delta t) = \langle Q(t) Q(t+\Delta t) \rangle.
\end{equation}
This ACF is not yet fully defined, because the choice of charge is not unique.
It can be that of the capacitor plates connected to the anode $Q_\textrm{a}$ or to the cathode $Q_\textrm{c}$ or the symmetrized charge $Q_\text{s} = (Q_\text{a}-Q_\text{c})/2$.
In a macroscopic capacitor, all three correlation functions would yield indistinguishable results, while for a finite system the one based on $Q_\textrm{s}$ yields a different result than the other two.
Those of the individual plates are indentical in our system, because the entire set-up is ideally symmetric.
%
Of course, the ACF of the symmetrized charge must be used to deduce the (quasi-) static capacitance of our circuit via the equipartition theorem
\begin{equation}
C = \frac{C_{Q Q}(\Delta t = 0)}{k_BT}.
\end{equation}

\begin{figure}[hbtp]
\vspace*{1mm}
\includegraphics[width=0.475\textwidth]{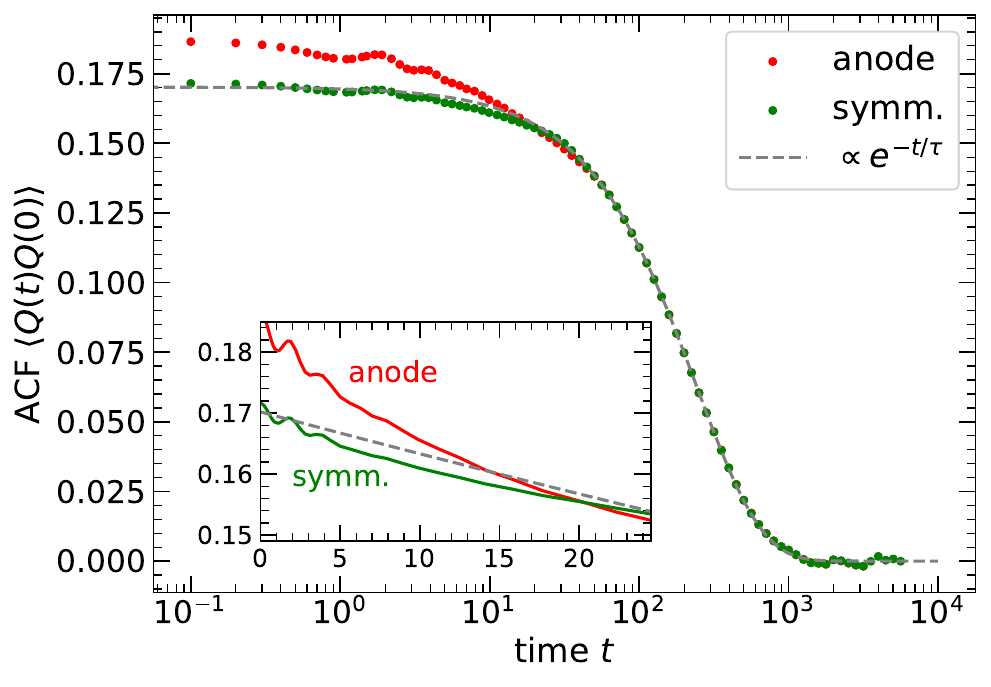}
\vspace*{-2mm}
\caption{\label{fig:nyquist}
Charge auto-correlation function (ACF), see Eq.~\eqref{eq:acf_q}, for the charge on one plate (anode), $Q_\text{a}$, and the symmetrized (symm.) charge $(Q_\text{a}-Q_\text{c})/2$, at a temperature of $T = 0.006$.
The dashed line is a fit to the data at $20<t<500$ to the ACF of the symmetrized charge with an exponential function. 
}
\end{figure}

Fig.~\ref{fig:nyquist} shows results deduced from a run with 100{,}000 time steps for equilibration and 64 million time steps, $\Delta t = 0.1$, for observation using the D-SQE method.
The fit to the ACF of the symmetrized charge on $30 < t < 500$ yields a value of $C(0) = 0.1702$ and a relaxation time of $\tau = 242$.
$C_{QQ}(0)$ can be converted in a capacitance of $C = C_{QQ}(0)/k_BT \to 28.37$.
This is within 4\% of the directly measured value.
The correlation time of $\tau = 242$ is even within 2.5\% of the relaxation time deduced earlier.
Thus, the thermal voltage of the 5,533 simulated split charges adds up to that of an individual large resistor.
Given the small deviation between 'continuum theory' and simulations, one could conclude that the continuum limit of text-book RC circuits would be an excellent description of  a 1,554-atom containing circuit, if electrons only behaved like a Fermi liquid in atom-thin wires.

\begin{figure}[hbtp]
\vspace*{1mm}
\includegraphics[width=0.475\textwidth]{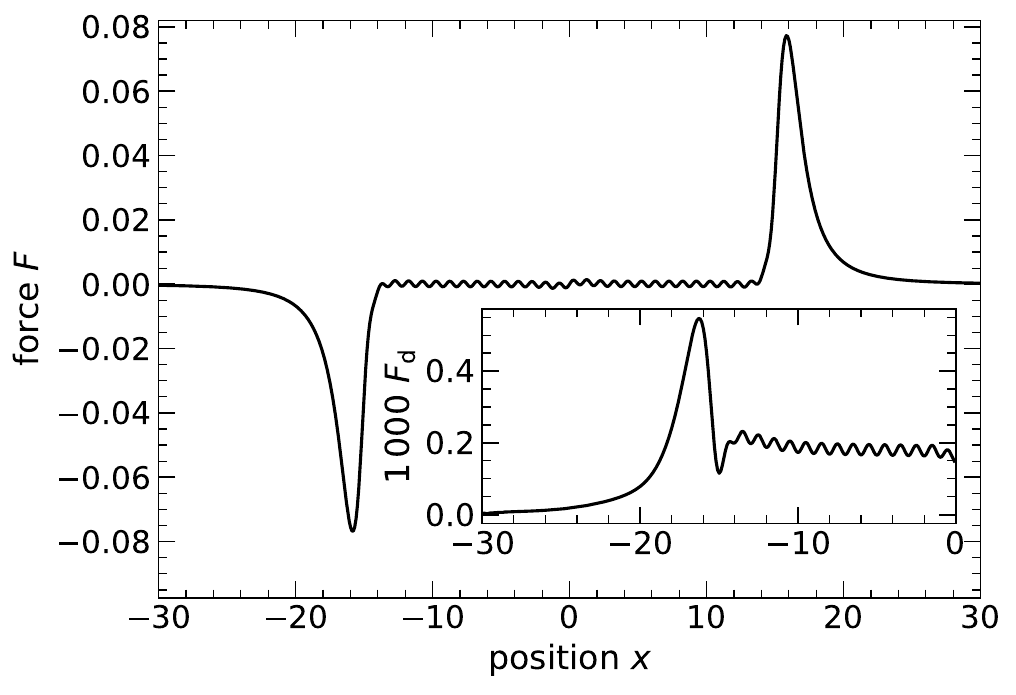}
\vspace*{-2mm}
\caption{\label{fig:force_sqe}
Main graph: Force on a unit charge moving at velocity $v = 0.01$ along the symmetry axis of a plane-parallel capacitor with a radius of $15~a_0$.
Inset: Damping force (scaled by a factor of 1{,}000) obtained by symmetrizing the force
$F_\text{d} = -(F(x)-F(-x))/2$.
}
\end{figure}

While the D-SQE model could be used to model response functions of regular dielectrics and metals, it is primarily meant as a tool to properly capture the dynamics of ions moving past polarizable but dissipative media. 
To show the feasibility of this application, the force acting on a unit charge moving at a constant velocity of $v = 0.01$ is computed and shown in Fig.~\ref{fig:force_sqe}.
The motion of the unit charge is confined to the radial direction of the capacitor as shown in the bottom panel of Fig.~\ref{fig:capaModel}, specifically midway between the two planes and parallel to the arrow.
The point charge experiences a large attractive force shortly before entering the capacitor, a rather small force while being between the plates and again an attractive force shortly after exiting.
In the capacitor, atomic discreteness of the two (commensurate) plates causes oscillations of the force with the period of the lattice. 
These oscillations are rather minor in particular in light of the distance between the charge and either plate being merely 3/2 times the lattice constant. 

Zooming into the main panel of Fig.~\ref{fig:force_sqe} near $x = 0$ reveals a small deviation from a (pseudo-) steady state.
This is mainly due to the interaction of the moving ion with the two front atoms of the wire being connected to capacitors.
The effect is clearly visible in movie M3, when the front atom of the upper wire briefly becomes blue when the patch indicating induced charge density is centered near the wire. 

Any conservative force between charge and capacitor must obey $F(x) = -F(-x)$ due to the symmetry of the set-up. 
However, the motion of the charge causes a dynamical response in the capacitor plates, leading to dissipation -- to some minor degree even without explicit Ohmic damping.
In the split-charge model, this would be because the split-charges in the wire-capacitor system will be in motion after a charge has traveled through them, in close analogy to the damping caused by atom/solid-surface scattering~\cite{Adelman1976JCP}.
The kinetic and potential energy contained in this motion will have to be provided by the (force acting on the) moving charge. 

The (frictional) work done on the system when moving between two symmetry-related points is $\int_{-x_0}^{x_0} \!\text{d}x F(x)$ so that the symmetrized force $F_\textrm{d}(x) = -(F(x)+F(-x))/2$ can be interpreted as an effective, instantaneous damping force as long as the moving charge's velocity is constrained to be constant. 
It is shown in the inset of Fig.~\ref{fig:force_sqe}.
Two contributions can be separated:
One that is due to the charging / decharging of the capacitor upon approach and retraction of the charge and another one due to the dragging of mirror charges through the capacitance. 

Although investigating the dissipation caused by the moving charge in detail would certainly be an interesting exercise, which might be done in some loose analogy to the modeling of viscous contact hysteresis, we restrict ourselves to a minimum analysis here with a single reference velocity. 
Reducing the velocity by a factor of two makes the damping $F_\textrm{d}(x)$ be almost exactly half of that of the reference calculation. 
If the resistivity is halfed, only the damping for $\vert x \vert \lesssim r$ is halfed, while the hysteretic loss related to approach and retraction is reduced less.
Thus, motion within the capacitor is impeded by Ohmic resistance, while the approach-retraction hysteresis has additional contributions. 

Given the results on the Stokes friction, the damping force satisfies $F_\text{d} \approx m \gamma v$ with $m \gamma \approx 0.02\to 4.74\cdot 10^{-17}$~N/s for the used parameterization and set-up. 
This would give a damping coefficient of $\gamma \approx 4.11$~GHz in case of a lithium ion and thus a slip time of 0.243~ns.
For comparison, the electronic damping coefficient of nitrogen sliding past a lead surface is merely 0.05~GHz~\cite{Dayo1998PRL}.

\section{Conclusions}

In this work, dissipative dyanmics were added to the split-charge model in a phenomenological fashion.
It allows the modeling of targeted dielectric properties in the framework of charge-equilibration processes by introducing Ohmic damping to the transfer of charge through a chemical bond.
However, the approach is not restricted to the atomic scale.
It can be defined in a way such that any linear-response, dielectric permittivity of a metal or a  dielectric medium can be reproduced at coarse scales.
The philosophy of the approach is similar to the Drude model.
However, the D-SQE model reflects the charging of heterogeneous or contacting solids in a natural way, while the Drude model would have to be generalized to reflect local work functions. 
Once the bond hardness is finite, the model is close in spirit to the Lorentz model.
However, the D-SQE model is meant to describe charge-transfer polarizability while the Lorentz model mimicks on-site polarizability. 
Of course, inducible point dipoles can be added to the model, as mentioned above. 
To alleviate computational burden, it may be advisable to use large pseudo-inductances, as this allows large time steps to be used.
This would impair the accuracy of modeling conductivity near optical frequencies, while the Ohmic response, which is typically the most relevant for molecular dynamics applications, would remain unaffected.

One shortcoming of the current formulation is that the Ohmic resistance is added in an \textit{ad-hoc} fashion.
In reality, Ohmic resistance has several origins, one of which is the scattering of phonons and electrons~\cite{ashcroft1976book}.
Such scattering can also occur naturally in the pursued approach, e.g., when the pseudo-inductance is a time dependent quantity due to fluctuating bond lengths.
Temporo-spatial variations of the pseudo-inductance densities would most naturally lead to the scattering of split-charge waves through ths system.
As long as ions and split charges are treated classicaly and the temperature is below the Debye temperature, that scattering should be small compared to the real one.
However, it \textit{might} be possible to emulate the phonon-electron scattering after quantizating the relevant degrees of freedom, for example, through Feynman's path-centroid density~\cite{Cao1994JCP}.
Such an approch has successfully reproduced damped dynamics of a commensurate, quantum mechanical Frenkel-Kontorova model, including the gap closure at sufficiently small mass~\cite{Krajewski2004PRL}.

Even without quantization, challenges remain for the D-SQE approach when used in regular potential or force-field based simulations, i.e., when the bonds related to a split charge have varying lengths and can break or form. 
Once a bond becomes long, 
$\kappa_\textrm{b}$ should increase and the pseudo-inductance (roughly linearly proportional) with it.
Such a scaling is not only physically meaningful, because electronic eigenfrequencies do not diverge when chemical bonds break, but it is also numerically desirable, as it does not require the time steps to be made extremely small. 
Nonetheless, the equation of motion would be a modified Langevin equation with an extra damping term proportional to $\dot{m}\dot{q}$. 
This might be a minor task compared to making force fields 'learn' realistic values for pseudo inductances and bond polarizabilites from calculations as those presented by Cheng and Verstraelen~\cite{Cheng2022JCP}


\bibliography{dsqe}

\end{document}